\documentclass[12pt]{iopart}
\newcommand{\bqa}{\begin{eqnarray}}
\newcommand{\eqa}{\end{eqnarray}}
\begin{document}

\title{Quantum Boltzman equation study for the Kondo breakdown
quantum critical point}
\author{K.-S. Kim$^{1}$ and C. P\'epin$^{2}$}

\address{$^{1}$Asia Pacific Center for Theoretical Physics,
Hogil Kim Memorial building 5th floor, POSTECH, Hyoja-dong, Namgu,
Pohang 790-784, Korea} \address{$^{2}$Institut de Physique
Th\'eorique, CEA, IPhT, CNRS, URA 2306, F-91191 Gif-sur-Yvette,
France} \ead{$^{1}$\mailto{kimks@apctp.org} and
$^{2}$\mailto{catherine.pepin@cea.fr}}

\begin{abstract}
We develop the quantum Boltzman equation approach for the Kondo
breakdown quantum critical point, involved with two bands for
conduction electrons and localized fermions. Particularly, the
role of vertex corrections in transport is addressed, crucial for
non-Fermi liquid transport of temperature linear dependence. Only
one band of spinons may be considered for scattering with gauge
fluctuations, and their associated vertex corrections are
introduced in the usual way, where divergence of self-energy
corrections is cancelled by that of vertex corrections, giving
rise to the physically meaningful result in the gauge invariant
expression for conductivity. On the other hand, two bands should
be taken into account for scattering with hybridization
excitations, giving rise to coupled quantum Boltzman equations. We
find that vertex corrections associated with hybridization
fluctuations turn out to be irrelevant due to heavy mass of
spinons in the so called decoupling limit, consistent with the
diagrammatic approach showing the non-Fermi liquid transport.
\end{abstract}



\maketitle

\section{Introduction}

Non-Fermi liquid transport phenomena near heavy fermion quantum
critical points (QCPs) are one of the central interests in the
business of strongly correlated electrons \cite{HFQCP}. To reveal
scattering mechanism between charge carriers and critical
fluctuations is the key to understand the temperature linear
resistivity \cite{TR}, one of the hallmarks for non-Fermi liquid
physics in the quantum critical matters. Basically, the nature of
critical modes, more concretely, the dynamical exponent $z$
determining their dispersion relation $\omega \sim q^{z}$ and the
scattering vertex between charge carriers and such critical modes
are essential ingredients for the transport phenomena.

The standard model of quantum criticality in a metallic system is
a $z = 2$ critical theory, often referred as Hertz-Moriya-Millis
(HMM) theory \cite{HMM}. Fortunately, many heavy fermion compounds
have been shown not to follow the spin-density-wave (SDW)
theoretical framework, where the temperature linear resistivity
\cite{LGW_F_QPT_Nature}, divergent Gr\"uneisen ratio with an
exponent $2/3$ \cite{GR_Exp}, Fermi surface reconstruction at the
QCP \cite{dHvA,Hall}, and the presence of localized magnetic
moments at the transition towards magnetism \cite{INS_Local_AF}
seem to support a more exotic scenario. An interesting suggestion
is that the heavy-fermion quantum transition is analogous to an
orbital selective Mott transition
\cite{DMFT,Senthil_Vojta_Sachdev,Pepin_KBQCP,Paul_KBQCP}, where
only the f-electrons experience the metal-insulator transition,
identified with a breakdown of the Kondo effect. This Kondo
breakdown scenario differs from the HMM theory \cite{HMM}, in
respect that the whole heavy Fermi surface is destabilized at the
QCP in the former case while only hot regions connected by SDW
vectors become unstable in the latter.

The nature of the Kondo breakdown QCP turns out to be multi-scale
\cite{Pepin_KBQCP,Paul_KBQCP}. Dynamics of hybridization
fluctuations is described by $z = 3$ critical theory due to Landau
damping of electron-spinon polarization above an intrinsic energy
scale $E^{*}$, while by $z = 2$ dilute Bose gas model below
$E^{*}$. The energy scale $E^{*}$ originates from mismatch of the
Fermi surfaces of conduction electrons and spinons, shown to vary
from ${\cal O}(10^{0})$ $mK$ to ${\cal O}(10^{2})$ $mK$. Based on
the $z = 3$ quantum criticality, both the logarithmic divergent
specific heat coefficient and power-law diverging thermal
expansion coefficient was shown to fit successfully, giving rise
to the divergent Gr\"uneisen ratio with an exponent $2/3$
\cite{Kim_GR}.

Recently, we showed that not only electrical resistivity but also
thermal resistivity shows quasi-linear temperature dependence
around the Kondo breakdown QCP due to scattering with $z = 3$
critical hybridization fluctuations \cite{Kim_TR}, based on the
Kubo formula where diagrammatic calculations were performed in the
$1/N$ expansion with the spin degeneracy $N$. An important point
is that vertex corrections for scattering with hybridization
fluctuations can be neglected, a unique feature of the two band
model, resulting from heavy mass of spinons
\cite{Pepin_KBQCP,Paul_KBQCP}. This allows us to replace the
transport time with the scattering time for such a process. Then,
only the self-energy correction will be involved for transport,
causing the non-Fermi liquid resistivity.

In this paper we clarify the issue related with vertex corrections
for the transport phenomena at the Kondo breakdown QCP, based on
quantum Boltzman equations where vertex corrections are introduced
naturally. An important feature is emergence of coupled quantum
Boltzman equations for the distribution function of each band. In
the heavy fermion phase described by condensation of Kondo bosons,
only the lowest heavy fermion band may be taken into account,
resulting in the Fermi liquid transport owing to the absence of
scattering with gapless fluctuations. In the fractionalized Fermi
liquid phase of the Kondo breakdown scenario
\cite{Senthil_Vojta_Sachdev}, hybridization fluctuations are
gapped, leaving the two bands decoupled in the low energy limit
and allowing us to consider the two quantum Boltzman equations
independently. On the other hand, at the Kondo breakdown QCP
critical hybridization fluctuations force the two quantum Boltzman
equations coupled, requiring to take both distribution functions
int account on equal footing. In this situation the quantum
Boltzman equation study was not preformed yet at least for the
heavy fermion QCP as far as we know.

The present paper is organized as follows. In section 2 we
introduce an effective field theory for the Kondo breakdown QCP.
In section 3 we examine the electrical resistivity based on the
coupled quantum Boltzman equations, where the quantum Boltzman
equation study is reviewed for the U(1) gauge theory of one band
model and its extension to the two band model is derived. In
section 4 we summarize our results.

\section{An effective field theory for the Kondo breakdown quantum
critical point}

We start from the Anderson lattice model in the large-$U$ limit
\bqa && L = \sum_{i} c_{i\sigma}^{\dagger}(\partial_{\tau} -
\mu)c_{i\sigma} - t \sum_{\langle ij \rangle}
(c_{i\sigma}^{\dagger}c_{j\sigma} + H.c.) \nonumber \\ && + V
\sum_{i} (d_{i\sigma}^{\dagger}c_{i\sigma} + H.c.) \nonumber \\ &&
+ \sum_{i}d_{i\sigma}^{\dagger}(\partial_{\tau} +
\epsilon_{f})d_{i\sigma} + J \sum_{\langle ij \rangle}
\vec{S}_{i}\cdot\vec{S}_{j} , \eqa where $c_{i\sigma}$ and
$d_{i\sigma}$ are conduction electron with a chemical potential
$\mu$ and localized electron with an energy level $\epsilon_{f}$.
The last spin-exchange term is introduced for competition with the
hybridization term of $V$.

Resorting to the U(1) slave-boson representation $d_{i\sigma} =
b_{i}^{\dagger} f_{i\sigma}$ with the single occupancy constraint
$b_{i}^{\dagger}b_{i} + f_{i\sigma}^{\dagger} f_{i\sigma} = S N$
to take strong correlations with $S = 1/2$, one can rewrite Eq.
(1) with \bqa && L = \sum_{i}
c_{i\sigma}^{\dagger}(\partial_{\tau} - \mu)c_{i\sigma} - t
\sum_{\langle ij \rangle} (c_{i\sigma}^{\dagger}c_{j\sigma} +
H.c.) \nonumber \\ && + V \sum_{i}
(b_{i}f_{i\sigma}^{\dagger}c_{i\sigma} +
H.c.) + \sum_{i}b_{i}^{\dagger} \partial_{\tau} b_{i} \nonumber \\
&& + \sum_{i}f_{i\sigma}^{\dagger}(\partial_{\tau} +
\epsilon_{f})f_{i\sigma} + \frac{J}{N} \sum_{\langle ij \rangle} (
f_{i\sigma}^{\dagger}\chi_{ij}f_{j\sigma} + H.c.) \nonumber \\ &&
+ i \sum_{i} \lambda_{i} (b_{i}^{\dagger}b_{i} +
f_{i\sigma}^{\dagger} f_{i\sigma} - 1) + \frac{J}{N} \sum_{\langle
ij \rangle} |\chi_{ij}|^{2} , \eqa where $b_{i}$ and $f_{i\sigma}$
are holon and spinon, associated with hybridization and spin
fluctuations, respectively. The spin-exchange term for the
localized orbital is decomposed via exchange hopping processes of
spinons, where $\chi_{ij}$ is a hopping parameter for the
decomposition. $\lambda_{i}$ is a Lagrange multiplier field to
impose the constraint, and $N$ is the number of fermion flavors
with $\sigma = 1, ..., N$.

Performing the saddle-point approximation of $b_{i} \rightarrow
b$, $\chi_{ij} \rightarrow \chi$, and $i\lambda_{i} \rightarrow
\lambda$, one finds an orbital selective Mott transition as
breakdown of the Kondo effect at $J \approx T_{K}$, where a
spin-liquid Mott insulator ($\langle b_{i} \rangle = 0$) arises in
$J > T_{K}$ while a heavy-fermion Fermi liquid ($\langle b_{i}
\rangle \not= 0$) results in $T_{K} > J$
\cite{Senthil_Vojta_Sachdev,Pepin_KBQCP,Paul_KBQCP}. Here, $T_{K}
= D \exp\Bigl(\frac{\epsilon_{f}}{N \rho_{c}V^{2}}\Bigr)$ is the
single impurity Kondo temperature, where $\rho_{c} \approx
(2D)^{-1}$ is the density of states for conduction electrons with
the half bandwidth $D$.

Quantum fluctuations should be incorporated for the critical
physics at the Kondo breakdown QCP, where two kinds of bosonic
collective modes will scatter two kinds of fermions, that is,
conduction electrons and spinons. Gauge fluctuations corresponding
to phase fluctuations of the hopping parameter $\chi_{ij} = \chi
e^{ia_{ij}}$ are introduced to express collective spin
fluctuations \cite{Nagaosa_Lee}. Hybridization fluctuations are
critical, playing an important role for the Kondo breakdown QCP.
Such four field variables lead us to the following effective field
theory in the continuum approximation, \bqa && {\cal L}_{ALM} =
c_{\sigma}^{*}(\partial_{\tau} - \mu_{c})c_{\sigma} +
\frac{1}{2m_{c}}|\partial_{i} c_{\sigma}|^{2} \nonumber \\  && +
f_{\sigma}^{*}(\partial_{\tau} - \mu_{f} - ia_{\tau})f_{\sigma} +
\frac{1}{2m_{f}}|(\partial_{i} - ia_{i})f_{\sigma}|^{2} \nonumber
\\ && +b^{*}(\partial_{\tau} - \mu_b -ia_{\tau})b +
\frac{1}{2m_{b}}|(\partial_{i} - ia_{i})b|^{2} + \frac{u_{b}}{2}
|b|^{4} \nonumber \\ &&+  V (b^{*}c_{\sigma}^{*}f_{\sigma} + H.c.)
\nonumber
\\ && + \frac{1}{4g^{2}} f_{\mu\nu}f_{\mu\nu} + S N (\mu_{b} + i
a_{\tau}) , \eqa where $g$ is an effective coupling constant
between matter and gauge fields, and several quantities, such as
fermion band masses and chemical potentials, are redefined as
follows \bqa & \lambda \rightarrow - \mu_{b} , ~~~ (2m_{c})^{-1} =
t , ~~~ (2m_{f})^{-1} = J \chi, \nonumber \\ & \mu_{c} = \mu + 2d
t , ~~~ - \mu_{f} = \epsilon_{f} +\lambda - 2 J d \chi . \nonumber
\eqa Fermion bare bands $\epsilon_{k}^{c}$ and $\epsilon_{k}^{f}$
for conduction electrons and spinons, respectively, are treated in
the continuum approximation as $\epsilon_{k}^{c} \approx - 2 d t +
t (k_{x}^{2} + k_{y}^{2} + k_{z}^{2})$ and $\epsilon_{k}^{f}
\approx - 2 J d \chi + J \chi (k_{x}^{2} + k_{y}^{2} +
k_{z}^{2})$. The band dispersion for hybridization can arise from
high energy fluctuations of conduction electrons and spinons.
Actually, the band mass of holons is given by $m_{b}^{-1} \approx
N V^{2} \rho_{c} /2$, where $\rho_{c}$ is the density of states
for conduction electrons \cite{Pepin_KBQCP,Paul_KBQCP}. Local
self-interactions denoted by $u_{b}$ can be introduced via
non-universal short-distance-scale physics. Maxwell dynamics for
gauge fluctuations appears from high energy fluctuations of
spinons and holons.

Based on the effective Lagrangian, recent studies
\cite{Pepin_KBQCP,Paul_KBQCP} developed an Eliashberg theory for
the Kondo breakdown QCP, where momentum dependence in fermion
self-energies and vertex corrections are neglected, allowing us to
evaluate one loop-level quantum corrections fully
self-consistently. Actually, this approximation was shown to be
"exact" in the large $N$ limit \cite{FMQCP}. The Eliashberg theory
for hybridization fluctuations results in the $z = 3$ Kondo
breakdown QCP, discussed in the introduction.

\section{Quantum Boltzman equation study}

We examine electrical transport at the Kondo breakdown QCP based
on quantum Boltzman equations, where we assume that both
hybridization and gauge fluctuations are in equilibrium and
consider only fermion contributions, consistent with the one-loop
result for the transport coefficient \cite{Kim_TR}. Since we have
two kinds of fermion excitations, we find coupled quantum Boltzman
equations for distributions of conduction electrons and spinons.
Solving such coupled quantum Boltzman equations, we find that the
diagrammatic result is recovered in the so called "decoupling"
limit of these equations, where vertex corrections for scattering
with hybridization fluctuations can be ignored, but those for
scattering with gauge fluctuations should be introduced in the
spinon conductivity.

Before we perform the quantum Boltzman equation study for the
Kondo breakdown QCP with two bands, we review on this approach in
the U(1) gauge theory with one band in order to understand the
role of vertex corrections in the transport coefficient
\cite{Nambu_Vertex,YB_Vertex} and demonstrate that our treatment
successfully recovers the known result \cite{YB_QBE,Nave_QBE}.

\subsection{Application to U(1) gauge theory for a spin liquid
state}

We apply the quantum Boltzman equation to the transport problem of
U(1) gauge theory, \bqa && S_{eff} = \int {d \tau} \int{d^{d}r}
\Bigl\{ \psi_{\sigma}^{\dagger}(\partial_{\tau} - i a_{\tau} -
\mu_{\psi})\psi_{\sigma} + \frac{1}{2m_{\psi}} |(
\partial_{i} - i a_{i}) \psi_{\sigma}|^{2} \Bigr\} \nonumber \\ && +
\int\frac{d \nu}{2\pi} \sum_{q} D(q,\nu) \Bigl(\delta_{ij} -
\frac{q_{i}q_{j}}{q^{2}}\Bigr) a_{i}(q,\nu) a_{j}(-q,-\nu) ,  \eqa
where $D(q,\nu) = \bigl(- i \gamma_{\psi} \frac{\nu}{q} +
\chi_{\psi} q^{2}\bigr)^{-1}$ is the gauge propagator with the
diamagnetic susceptibility $\chi_{\psi}$ and Landau damping
coefficient $\gamma_{\psi}$.

One can obtain this effective field theory from the Hubbard model
in the frustrated lattice based on the U(1) slave-rotor
representation \cite{U1SR}, where charge fluctuations are gapped
at half filling, but magnetic ordering is prohibited owing to the
geometrical frustration, corresponding to a spin liquid Mott
insulator. One also finds this effective field theory in the so
called algebraic charge liquid for the anomalous normal state of
high T$_{c}$ cuprates, derived from the U(1) slave-fermion
representation, where spin fluctuations described by Schwinger
bosons are gapped, but charged excitations represented by
fermionic holons are gapless, allowing an anomalous metallic state
due to scattering with gauge fluctuations \cite{ACL,SF_Kim}.

Compared with the effective field theory for the Kondo breakdown
QCP of the Anderson lattice model, this U(1) gauge theory is a
simplified version since it does not have both holons and
conduction electrons. In this section we focus on the mathematical
structure, in particular, the gauge invariant expression for
conductivity \cite{YB_QBE,Nave_QBE} instead of the physical
aspect, in order to prepare for the Boltzman equation study of the
Anderson lattice model.

We start from the quantum Boltzman equation \cite{Mahan_QBE} \bqa
&& [\partial_{\omega} f(\omega)] \Gamma(k,\omega)
[A(k,\omega)]^{2} \mathbf{v}_{k} \cdot \mathbf{E} =
I_{coll}(k,\omega) , \eqa where $\Gamma(k,\omega)$ and
$A(k,\omega)$ are the imaginary parts of the retarded self-energy
and retarded Green's function, respectively, $f(\omega)$ is the
Fermi-Dirac distribution function in equilibrium, $\mathbf{v}_{k}$
is the velocity of fermions, and $\mathbf{E}$ is an external
electric field. $I_{coll}(k,\omega)$ is the collision term given
by \bqa && I_{coll}(k,\omega) =
\Sigma^{>}(k,\omega)G^{<}(k,\omega) -
\Sigma^{<}(k,\omega)G^{>}(k,\omega) ,  \eqa where
$\Sigma^{<,>}(k,\omega)$ and $G^{<,>}(k,\omega)$ are lesser and
greater self-energies and Green's functions, respectively. Using
the identity of \bqa && \Sigma^{>}(k,\omega)G^{<}(k,\omega) -
\Sigma^{<}(k,\omega)G^{>}(k,\omega)  \nonumber \\ && = 2i
\Gamma(k,\omega) G^{<}(k,\omega) - i\Sigma^{<}(k,\omega)
A(k,\omega) \nonumber , \eqa where the lesser self-energy is given
by \bqa && \Sigma^{<}(k,\omega) = \sum_{q} \int_{0}^{\infty}
\frac{d\nu}{\pi}
\Bigl| \frac{k\times\hat{q}}{m_{\psi}} \Bigr|^{2} \Im D(q,\nu) \nonumber \\
&& [\{n(\nu) + 1\} G^{<}(k+q,\omega+\nu) + n(\nu)
G^{<}(k+q,\omega-\nu)]     \eqa with the Bose-Einstein
distribution function $n(\nu)$ in the one loop approximation, the
lesser Green's function is the only unknown function, determined
by the quantum Boltzman equation. This quantum Boltzman equation
is well derived in Ref. \cite{Mahan_QBE}, based on the
Schwinger-Keldysh formulation.

In the linear response regime we can expand the lesser Green's
function up to the first order of an electric field \bqa &&
G^{<}(k,\omega) = i A(k,\omega) \Bigl[ f(\omega) - \Bigl(
\frac{\partial f(\omega)}{\partial \omega} \Bigr) \mathbf{E} \cdot
\mathbf{v}_{k} \Lambda(k,\omega) \Bigr] ,   \eqa where
$\Lambda(k,\omega)$ is the distribution function out of, but near
equilibrium due to the electric field. Inserting this ansatz into
the lesser self-energy, we obtain the following expression for the
lesser self-energy \bqa && \Sigma^{<}(k,\omega) = i \sum_{q}
\int_{0}^{\infty} \frac{d\nu}{\pi} \Bigl|
\frac{k\times\hat{q}}{m_{\psi}} \Bigr|^{2} \Im D(q,\nu) f(\omega)
\nonumber
\\ && \Bigl\{ [ n(\nu) + f(\omega+\nu) ] A(k+q,\omega+\nu)
- [ n(-\nu) + f(\omega-\nu) ] A(k+q,\omega-\nu) \Bigr\} \nonumber
\\ && + i \sum_{q} \int_{0}^{\infty} \frac{d\nu}{\pi} \Bigl|
\frac{k\times\hat{q}}{m_{\psi}} \Bigr|^{2} \Im D(q,\nu) \mathbf{E}
\cdot \mathbf{v}_{k+q} \Bigl( -\frac{\partial f(\omega)}{\partial
\omega} \Bigr) \nonumber \\ && \Bigl\{ [ n(\nu) + f(\omega+\nu) ]
\frac{1 - f(\omega+\nu)}{1 - f(\omega)} A(k+q,\omega+\nu)
\Lambda(k+q,\omega+\nu) \nonumber \\ && - [ n(-\nu) +
f(\omega-\nu) ] \frac{1 - f(\omega-\nu)}{1 - f(\omega)}
A(k+q,\omega-\nu) \Lambda(k+q,\omega-\nu) \Bigr\} , \eqa where we
used the identities for thermal factors of fermions and bosons,
\bqa && \{n(\nu) + 1\} f(\omega+\nu) = f(\omega) \{ n(\nu) +
f(\omega+\nu) \} , \nonumber \\ && n(\nu) f(\omega - \nu) = -
f(\omega) \{ n(-\nu) + f(\omega-\nu) \} , \nonumber \eqa and \bqa
&& \{n(\nu) + 1\} \Bigl( -\frac{\partial f(\omega+\nu)}{\partial
\omega} \Bigr) = \{ n(\nu) + f(\omega+\nu) \}  \frac{1 -
f(\omega+\nu)}{1 - f(\omega)} \Bigl( -\frac{\partial
f(\omega)}{\partial \omega} \Bigr) ,  \nonumber \\ && n(\nu)
\Bigl( -\frac{\partial f(\omega-\nu)}{\partial \omega} \Bigr)
\nonumber = - \{ n(-\nu) + f(\omega-\nu) \}  \frac{1 -
f(\omega-\nu)}{1 - f(\omega)} \Bigl( -\frac{\partial
f(\omega)}{\partial \omega} \Bigr) . \nonumber \eqa

Inserting both the lesser Green's function and self-energy into
the quantum Boltzman equation, we find \bqa &&
\Lambda(k_{F},\omega) \approx \frac{1}{2} A(k_{F},\omega) +
\frac{1}{ 2\Gamma(k_{F},\omega) } \sum_{q} \int_{0}^{\infty}
\frac{d\nu}{\pi} \Bigl| \frac{k_{F}\times\hat{q}}{m_{\psi}}
\Bigr|^{2} \Im D(q,\nu) \nonumber \\ && \Bigl\{ [ n(\nu) +
f(\omega+\nu) ] A(k_{F}+q,\omega+\nu) - [ n(-\nu) + f(\omega-\nu)
] A(k_{F}+q,\omega-\nu) \Bigr\} \nonumber \\ && \Bigl(
\frac{\mathbf{v}_{k_{F}}\cdot\mathbf{v}_{k_{F}+q}}{v_{F}^{2}}
\Bigr) \Lambda(k_{F},\omega) , \eqa where the momentum is replaced
with the Fermi momentum $k_{F}$ because usual transport phenomena
occur near the Fermi surface except some topological quantities
such as Hall conductivity \cite{Hall_Conductivity} and frequency
dependence in both the "vertex-distribution" function
$\Lambda(k_{F},\omega)$ and thermal Fermi factor is simplified. In
this expression the imaginary part of the self-energy or
scattering rate is defined as \bqa && 2 \Gamma(k,\omega) =
\sum_{q} \int_{0}^{\infty} \frac{d\nu}{\pi} \Bigl|
\frac{k\times\hat{q}}{m_{\psi}} \Bigr|^{2} \Im D(q,\nu) \Bigl\{ [
n(\nu) + f(\omega+\nu) ] A(k+q,\omega+\nu) \nonumber
\\ && - [ n(-\nu) +
f(\omega-\nu) ] A(k+q,\omega-\nu) \Bigr\} . \eqa This
approximation will be justified by the fact that it gives rise to
the known result in the gauge theory context.

Introducing the relative angle $\theta$ between the initial
$k_{F}$ and final $k_{F} + q$ momenta, we obtain \bqa &&
\Lambda(k_{F},\omega) =
\frac{2\Gamma(k_{F},\omega)}{2\Gamma_{1-\cos}(k_{F},\omega)}
A(k_{F},\omega) , \eqa where \bqa &&
2\Gamma_{1-\cos}(k_{F},\omega) = \frac{3}{2\Lambda^{3}}
\int_{0}^{\Lambda} d q q^{2} \int_{-1}^{1} {d\cos\theta}
[v_{F}^{2}\cos^{2}(\theta/2)] \nonumber \\ && \int_{0}^{\infty}
\frac{d\nu}{\pi} \Im D(q,\nu) [1 - \cos\theta] \Bigl\{ [ n(\nu) +
f(\omega+\nu) ] A(k_{F}+q,\omega+\nu) \nonumber \\ && - [ n(-\nu)
+ f(\omega-\nu) ] A(k_{F}+q,\omega-\nu) \Bigr\} . \eqa In this
expression $\sum_{q}$ is replaced with $\frac{3}{2\Lambda^{3}}
\int_{0}^{\Lambda} d q q^{2} \int_{-1}^{1} {d\cos\theta}$ in $d =
3$, where $\Lambda$ is a momentum cutoff. $1 - \cos \theta$ factor
in $\Gamma_{1-\cos}(k_{F},\omega)$ identifies
$[2\Gamma_{1-\cos}(k_{F},\omega)]^{-1}$ with the transport time
$\tau_{tr}(\omega)$, capturing large angle scattering dominantly.

The electrical (charge) or number conductivity is expressed by the
lesser Green's function, \bqa && J_{\mu}^{\psi} = - i
\int\frac{d^{3}k}{(2\pi)^{3}} \frac{k_{\mu}}{m_{\psi}}
\int\frac{d\omega}{2\pi} G^{<}(k,\omega) . \eqa Inserting the
near-equilibrium ansatz for the lesser Green's function into this
expression, we obtain the electrical conductivity \bqa
\sigma_{\mu\nu}(T) && = \int\frac{d^{3}k}{(2\pi)^{3}}
\int\frac{d\omega}{2\pi} v_{k\mu}v_{k\nu} \Bigl( - \frac{\partial
f(\omega)}{\partial \omega} \Bigr) A(k,\omega) \Lambda(k,\omega)
\eqa where the equilibrium contribution does not generate
currents, thus vanishes.

Inserting the vertex-distribution function into the conductivity
expression and performing the momentum and energy integration, we
reach the final form of the conductivity \bqa && \sigma(T) \approx
\mathcal{C} N_{F} v_{F}^{2} \tau_{tr}(T) \eqa with $\mathcal{C} =
\frac{N}{\pi} \int_{-\infty}^{\infty}{d y}
\frac{1}{(y^{2}+1)^{2}}$, where $N_{F}$ is the density of states
at the Fermi surface and the transport time is $\tau_{tr}(T) =
[2\Gamma_{1-\cos}(T)]^{-1}$, as emphasized before.

The transport time turns out to be $\tau_{tr}(T) \propto
T^{-5/3}$, giving rise to $\sigma(T) \propto T^{-5/3}$ in $d = 3$,
completely consistent with the previous study
\cite{YB_QBE,Nave_QBE}. An important point is that although the
self-energy correction due to gauge fluctuations is diverging at
finite temperatures, the gauge invariant expression for the
conductivity allows only the finite result, cancelling the
divergence via the vertex correction \cite{YB_Vertex}. $1 -
\cos\theta$ guarantees such cancellation. This is the power of the
quantum Boltzman equation, imposing the vertex correction
naturally.

In this derivation gauge fluctuations are assumed to be in
equilibrium. Generally speaking, their non-equilibrium
distribution due to external fields should be introduced.
Actually, phonon drag effects are well known in the
electron-phonon system \cite{Mahan_QBE}. Recently, this issue was
considered in the spin liquid context with $z = 3$ gauge
fluctuations \cite{Nave_QBE}, where coupled quantum Boltzmann
equations for spinon and photon distribution functions are
derived. It was argued that such coupled transport equations can
be decoupled in some cases, where such drag effects are
subdominant, compared with fermion contributions.

The present formulation differs from the previous approach in the
fact that we did not decompose the gauge field as the study of
Refs. \cite{YB_QBE,Nave_QBE}, where the low energy gauge field
giving rise to divergence is neglected and only high energy gauge
fluctuations are taken. Although the vertex-distribution function
itself is not well defined because its part corresponding to the
scattering rate is divergent at finite temperatures, we found that
such decomposition is not necessary because the formal divergence
should be cancelled in the last gauge invariant physical
expression. This spirit goes exactly through that of the
diagrammatic study.

\subsection{Application to the Kondo breakdown QCP of the Anderson lattice model}

In the Kondo breakdown scenario we have four kinds of field
variables, corresponding to conduction electrons, spinons, holons
(hybridization fluctuations), and gauge bosons (collective spin
fluctuations). Our main assumption for the transport study based
on the quantum Boltzman equation approach is that both
hybridization and gauge fluctuations are in equilibrium, as
pointed out earlier. This assumption is justified by the
diagrammatic study \cite{Kim_TR}, where contributions from boson
excitations are much smaller than fermion contributions, and by
the Boltzman equation study of the U(1) gauge theory discussed in
the previous section. As a result, we are allowed to have two
coupled quantum Boltzman equations, \bqa && [A_{c}(k,\omega)]^{2}
\partial_{\omega} f(\omega) \mathbf{E} \cdot \mathbf{v}_{k}^{c}
\Gamma_{c}(k,\omega) = I_{coll}^{c}(k,\omega) \nonumber , \\ &&
I_{coll}^{c}(k,\omega) = 2i \Gamma_{c}(k,\omega)
G^{<}_{c}(k,\omega) - i \Sigma^{<}_{c}(k,\omega) A_{c}(k,\omega)
\eqa for conduction electrons and \bqa && [A_{f}(k,\omega)]^{2}
\partial_{\omega} f(\omega) \mathbf{E} \cdot \mathbf{v}_{k}^{f}
\Gamma_{f}(k,\omega) = I_{coll}^{f}(k,\omega) \nonumber , \\ &&
I_{coll}^{f}(k,\omega) = 2i \Gamma_{f}(k,\omega)
G^{<}_{f}(k,\omega) - i \Sigma^{<}_{f}(k,\omega) A_{f}(k,\omega)
\eqa  for spinons.

\subsubsection{Contribution of conduction electrons}

The lesser self-energy for conduction electrons arises from
scattering with hybridization fluctuations, given by \bqa &&
\Sigma^{<}_{c}(k,\omega) = V^{2} \sum_{q} \int_{0}^{\infty}
\frac{d\nu}{\pi} \Im D_{b}(q,\nu) \nonumber \\ && [\{n(\nu) + 1\}
G^{<}_{f}(k+q,\omega+\nu) + n(\nu) G^{<}_{f}(k+q,\omega-\nu)]
\eqa in the Eliashberg framework. Since the spinon Green's
function appears in the electron self-energy, the two quantum
Boltzman equations are coupled with each other. This coupling
effect is the main character for the quantum Boltzman equation of
the Anderson lattice model at the QCP.

Inserting the lesser Green's function of spinons \bqa &&
G^{<}_{f}(k,\omega) = i A_{f}(k,\omega) \Bigl[ f(\omega) - \Bigl(
\frac{\partial f(\omega)}{\partial \omega} \Bigr) \mathbf{E} \cdot
\mathbf{v}_{k}^{f} \Lambda_{f}(k,\omega) \Bigr]  \eqa into the
electron lesser self-energy and the lesser Green's function for
conduction electrons \bqa && G^{<}_{c}(k,\omega) = i
A_{c}(k,\omega) \Bigl[ f(\omega) - \Bigl( \frac{\partial
f(\omega)}{\partial \omega} \Bigr) \mathbf{E} \cdot
\mathbf{v}_{k}^{c} \Lambda_{c}(k,\omega) \Bigr]   \eqa into the
quantum Boltzman equation for conduction electrons, we obtain \bqa
&& \Lambda_{c}(k_{F}^{c},\omega) \approx \frac{1}{2}
A_{c}(k_{F}^{c},\omega) + \frac{V^{2}}{
2\Gamma_{c}(k_{F}^{c},\omega) } \sum_{q} \int_{0}^{\infty}
\frac{d\nu}{\pi} \Im D_{b}(q,\nu) \Bigl( \frac{
\mathbf{v}_{k+q}^{f}\cdot\mathbf{v}_{k}^{c} }{v_{k}^{c2}} \Bigr)
\nonumber \\ && \Bigl\{ [ n(\nu) + f(\omega+\nu) ]
A_{f}(k_{F}^{f}+q,\omega+\nu) \nonumber \\ && - [ n(-\nu) +
f(\omega-\nu) ] A_{f}(k_{F}^{f}+q,\omega-\nu) \Bigr\}
\Lambda_{f}(k_{F}^{f},\omega) ,  \eqa where \bqa && 2
\Gamma_{c}(k,\omega) = V^{2} \sum_{q} \int_{0}^{\infty}
\frac{d\nu}{\pi} \Im D_{b}(q,\nu) \Bigl\{ [ n(\nu) + f(\omega+\nu)
] \nonumber \\ && A_{f}(k+q,\omega+\nu) - [ n(-\nu) +
f(\omega-\nu) ] A_{f}(k+q,\omega-\nu) \Bigr\}     \eqa is the
scattering rate of conduction electrons and the same
approximations as the case of the U(1) gauge theory are utilized.
It is important to notice that the vertex-distribution function
for conduction electrons is related with that for spinons. We
should know the vertex-distribution function for spinons.

\subsubsection{Contribution of spinons}

The lesser self-energy for spionon excitations results from
scattering with both hybridization and gauge fluctuations, given
by \bqa && \Sigma_{f}^{<}(k,\omega) = \Sigma_{f}^{b<}(k,\omega) +
\Sigma_{f}^{a<}(k,\omega) , \nonumber \\ &&
\Sigma^{b<}_{f}(k,\omega) = V^{2} \sum_{q} \int_{0}^{\infty}
\frac{d\nu}{\pi} \Im D_{b}(q,\nu) \nonumber
\\ && [\{n(\nu) + 1\} G^{<}_{c}(k+q,\omega+\nu) + n(\nu)
G^{<}_{c}(k+q,\omega-\nu)] , \nonumber \\ && \Sigma_{f}^{a<}
(k,\omega) = \sum_{q} \int_{0}^{\infty} \frac{d\nu}{\pi} \Bigl|
\frac{k\times\hat{q}}{m_{f}} \Bigr|^{2} \Im D_{a}(q,\nu) \nonumber
\\ && [\{n(\nu) + 1\} G^{<}_{f}(k+q,\omega+\nu) + n(\nu)
G^{<}_{f}(k+q,\omega-\nu)] ,     \eqa where the lesser Green's
function of conduction electrons appear in the
hybridization-vertex-induced spinon self-energy while that of
spinons arises in the self-energy correction via gauge
fluctuations.

Inserting the lesser Green's functions for both conduction
electrons and spinons into the lesser self-energy and quantum
Boltzman equation for spinons, we find \bqa && \mathbf{v}_{F}^{f}
\Lambda_{f}(k_{F}^{f},\omega) \approx \frac{\mathbf{v}_{F}^{f}}{2}
A_{f}(k_{F}^{f},\omega) + \frac{V^{2}}{ 2
\Gamma_{f}(k_{F}^{f},\omega)} \sum_{q} \int_{0}^{\infty}
\frac{d\nu}{\pi} \Im D_{b}(q,\nu) \mathbf{v}_{k_{F}+q}^{c}
\nonumber \\ && \Bigl\{ [ n(\nu) + f(\omega+\nu) ]
A_{c}(k_{F}^{c}+q,\omega+\nu) \nonumber
\\ && - [ n(-\nu) + f(\omega-\nu) ] A_{c}(k_{F}^{c}+q,\omega-\nu)
\Bigr\} \Lambda_{c}(k_{F}^{c} ,\omega ) \nonumber \\ && +
\frac{1}{ 2 \Gamma_{f}(k_{F}^{f},\omega)} \sum_{q}
\int_{0}^{\infty} \frac{d\nu}{\pi} \Bigl|
\frac{k_{F}^{f}\times\hat{q}}{m_{f}} \Bigr|^{2} \Im D_{a}(q,\nu)
\mathbf{v}^{f}_{k_{F}+q} \nonumber \\ && \Bigl\{ [ n(\nu) +
f(\omega+\nu) ] A_{f}(k_{F}^{f}+q,\omega+\nu) \nonumber \\ && - [
n(-\nu) + f(\omega-\nu) ] A_{f}(k_{F}^{f}+q,\omega-\nu) \Bigr\}
\Lambda_{f}(k_{F}^{f},\omega )  ,  \eqa where \bqa && 2
\Gamma_{f}(k,\omega) = V^{2} \sum_{q} \int_{0}^{\infty}
\frac{d\nu}{\pi} \Im D_{b}(q,\nu) \Bigl\{ [ n(\nu) + f(\omega+\nu)
] A_{c}(k+q,\omega+\nu) \nonumber \\ && - [ n(-\nu) +
f(\omega-\nu) ] A_{c}(k+q,\omega-\nu) \Bigr\} \nonumber \\ && +
\sum_{q} \int_{0}^{\infty} \frac{d\nu}{\pi} \Bigl|
\frac{k\times\hat{q}}{m_{f}} \Bigr|^{2} \Im D_{a}(q,\nu) \Bigl\{ [
n(\nu) + f(\omega+\nu) ] A_{f}(k+q,\omega+\nu) \nonumber \\ && - [
n(-\nu) + f(\omega-\nu) ] A_{f}(k+q,\omega-\nu) \Bigr\} \nonumber
\\ && \equiv 2 \Gamma_{f}^{b}(k,\omega) + 2
\Gamma_{f}^{a}(k,\omega) \eqa is the scattering rate of spinons
resulting from scattering with both hybridization
$\Gamma_{f}^{b}(k,\omega)$ and gauge fluctuations
$\Gamma_{f}^{a}(k,\omega)$. One can check Eqs. (25) and (26),
considering that the hybridization-induced part is basically the
same as that of the quantum Boltzman equation for conduction
electrons and the gauge-fluctuation part coincides with that shown
in the U(1) gauge theory of the previous section.

Inserting the vertex-distribution function [Eq. (22)] for
conduction electrons into the vertex-distribution function [Eq.
(25)] for spinons, we find the following expression for spinons
\bqa && \Lambda_{f}(k_{F}^{f},\omega) = \frac{1}{2} \Bigl\{
A_{f}(k_{F}^{f},\omega) +
\frac{\Gamma_{f,\cos}^{b}(k_{F}^{f},\omega) }{
\Gamma_{f}(k_{F}^{f},\omega)} A_{c}(k_{F}^{c},\omega) \Bigr\} \nonumber \\
&& + \Bigl\{ \frac{\Gamma_{f,\cos}^{b}(k_{F}^{f},\omega) }{
\Gamma_{f}(k_{F}^{f},\omega)}
\frac{\Gamma_{c,\cos}(k_{F}^{f},\omega)}{
\Gamma_{c}(k_{F}^{c},\omega) } +
\frac{\Gamma_{f,\cos}^{a}(k_{F}^{f},\omega) }{
\Gamma_{f}(k_{F}^{f},\omega)} \Bigr\}
\Lambda_{f}(k_{F}^{f},\omega)    \eqa
with \bqa && 2 \Gamma_{f,\cos}^{b}(k_{F}^{f},\omega) \equiv V^{2}
\frac{3}{2\Lambda^{3}} \int_{0}^{\Lambda} d q q^{2} \int_{-1}^{1}
d\cos\theta_{cf} \Bigl( \frac{v_{F}^{c}}{v_{F}^{f}} \cos
\theta_{cf} \Bigr) \nonumber \\ && \int_{0}^{\infty}
\frac{d\nu}{\pi} \Im D_{b}(q,\nu) \Bigl\{ [ n(\nu) + f(\omega+\nu)
] A_{c}(k_{F}^{c}+q,\omega+\nu) \nonumber \\ && - [ n(-\nu) +
f(\omega-\nu) ] A_{c}(k_{F}^{c}+q,\omega-\nu) \Bigr\} , \nonumber
\\ && 2 \Gamma_{f,\cos}^{a}(k_{F}^{f},\omega) \equiv
\frac{3}{2\Lambda^{3}} \int_{0}^{\Lambda} d q q^{2} \int_{-1}^{1}
d\cos\theta_{ff} \cos \theta_{ff} \nonumber \\ &&
\int_{0}^{\infty} \frac{d\nu}{\pi} [v_{F}^{f2}
\cos^{2}(\theta_{ff}/2)] \Im D_{a}(q,\nu) \Bigl\{ [ n(\nu) +
f(\omega+\nu) ] \nonumber \\ && A_{f}(k_{F}^{f}+q,\omega+\nu) - [
n(-\nu) + f(\omega-\nu) ] A_{f}(k_{F}^{f}+q,\omega-\nu) \Bigr\} ,
\nonumber
\\ && 2 \Gamma_{c,\cos}(k_{F}^{f},\omega) \equiv V^{2}
\frac{3}{2\Lambda^{3}} \int_{0}^{\Lambda} d q q^{2} \int_{-1}^{1}
d\cos\theta_{cf} \Bigl( \frac{v_{F}^{f}} {v_{F}^{c}}
\cos\theta_{cf} \Bigr) \nonumber \\ && \int_{0}^{\infty}
\frac{d\nu}{\pi} \Im D_{b}(q,\nu) \Bigl\{ [ n(\nu) + f(\omega+\nu)
] A_{f}(k_{F}^{f}+q,\omega+\nu) \nonumber \\ && - [ n(-\nu) +
f(\omega-\nu) ] A_{f}(k_{F}^{f}+q,\omega-\nu) \Bigr\} , \eqa where
$\theta_{cf}$ represents an angle between the initial electron
velocity $v_{F}^{c}$ and final spinon velocity $v_{F}^{f}$ and
$\theta_{ff}$ is defined in the similar way, but between spinons.

We obtain the spinon vertex-distribution function  \bqa &&
\Lambda_{f}(k_{F}^{f},\omega) = \frac{1}{2} \Bigl\{
\Gamma_{f}(k_{F}^{f},\omega)A_{f}(k_{F}^{f},\omega) +
\Gamma_{f,\cos}^{b}(k_{F}^{f},\omega) A_{c}(k_{F}^{c},\omega)
\Bigr\} \nonumber \\ && \Bigl\{\Gamma_{f}^{b}(k_{F}^{f},\omega) +
\Gamma_{f,1-\cos}^{a}(k_{F}^{f},\omega) -
\Gamma_{f,\cos}^{b}(k_{F}^{f},\omega)
\frac{\Gamma_{c,\cos}(k_{F}^{f},\omega)}{
\Gamma_{c}(k_{F}^{c},\omega) } \Bigr\}^{-1}   , \eqa where \bqa &&
2 \Gamma_{f,1-\cos}^{a}(k_{F}^{f},\omega) \equiv
\frac{3}{2\Lambda^{3}} \int_{0}^{\Lambda} d q q^{2} \int_{-1}^{1}
d\cos\theta_{ff} [1-\cos \theta_{ff}] \nonumber \\ &&
\int_{0}^{\infty} \frac{d\nu}{\pi} [v_{F}^{f2}
\cos^{2}(\theta_{ff}/2)] \Im D_{a}(q,\nu) \Bigl\{ [ n(\nu) +
f(\omega+\nu) ] A_{f}(k_{F}^{f}+q,\omega+\nu) \nonumber \\ && - [
n(-\nu) + f(\omega-\nu) ] A_{f}(k_{F}^{f}+q,\omega-\nu) \Bigr\}
\eqa is identified with $[\tau_{tr}^{a}(\omega)]^{-1}$ as shown in
the U(1) gauge theory of the previous section.

\subsubsection{Conductivity in the decoupling limit}

In the vertex-distribution function for spinons [Eq. (25)] we
neglect the coupling term $\Lambda_{c}(k_{F}^{c},\omega)$ as the
zeroth order approximation for the transport study, named as the
decoupling limit. One may understand validity of this
approximation, based on the fact that spinons are heavily massive
denoted by $\alpha \ll 1$ and scattering with conduction electrons
will not affect their dynamics much. Then, we find \bqa &&
\Lambda_{f}(k_{F}^{f},\omega) = \frac{1}{2} \frac{
\Gamma_{f}(k_{F}^{f},\omega)A_{f}(k_{F}^{f},\omega)
}{\Gamma_{f}^{b}(k_{F}^{f},\omega) +
\Gamma_{f,1-\cos}^{a}(k_{F}^{f},\omega) }  . \eqa

Inserting this expression into the spinon conductivity, we obtain
\bqa && \sigma_{f}(T) = \frac{\mathcal{C} N_{F}^{f}
v_{F}^{f2}}{2\Gamma_{f}^{b}(T) + 2\Gamma_{f,1-\cos}^{a}(T)} , \eqa
exactly the same as that of the diagrammatic study \cite{Kim_TR},
where vertex corrections are introduced only for the scattering
channel with gauge fluctuations. As discussed in the previous
section, $z = 3$ gauge fluctuations give rise to divergence for
self-energy corrections to spinons, but cancelled by vertex
corrections, allowing the gauge invariant finite physical
conductivity proportional to $\sim T^{-5/3}$ in $d = 3$. One may
ask why the same situation does not happen for scattering with $z
= 3$ hybridization fluctuations. Actually, such $z = 3$ dynamics
of holons is cut by an intrinsic energy scale $E^{*}$, and
scattering with $z = 2$ hybridization fluctuations below $E^{*}$
does not cause the divergence for self-energy corrections.

The vertex-distribution function for conduction electrons becomes
\bqa && \Lambda_{c}(k_{F}^{c},\omega) \approx \frac{1}{2}
A_{c}(k_{F}^{c},\omega) + \frac{1}{2}
\frac{\Gamma_{c,\cos}(k_{F}^{c},\omega)}{
\Gamma_{c}(k_{F}^{c},\omega) } \frac{
\Gamma_{f}(k_{F}^{f},\omega)A_{f}(k_{F}^{f},\omega)
}{\Gamma_{f}^{b}(k_{F}^{f},\omega) +
\Gamma_{f,1-\cos}^{a}(k_{F}^{f},\omega) } , \eqa where scattering
with spinons is incorporated through the vertex-distribution
function for spinons because light conduction electrons can be
much affected. However, calling \bqa &&
\frac{\Gamma_{c,\cos}(k_{F}^{c},\omega)}{
\Gamma_{c}(k_{F}^{c},\omega)} = \mathcal{O}(v_{F}^{f}/v_{F}^{c})
\approx \alpha \ll 1 , \eqa the second contribution in the
electron vertex-distribution can be neglected. As a result, the
conductivity from conduction electrons is free from vertex
corrections, becoming \bqa && \sigma_{c}(T) = \mathcal{C}
\frac{N_{F}^{c} v_{F}^{c2}}{2\Gamma_{c}(T)} , \eqa which coincides
with that of the diagrammatic study \cite{Kim_TR} showing
$\Gamma_{c}(T) \sim T \ln (T/E^{*})$ in the $z = 3$ critical
regime.

The last work is to find an actual expression for the physical
conductivity, referred as the Ioffe-Larkin composition rule
\cite{IF_Cond} \bqa && \sigma(T) = \sigma_{c}(T) +
\frac{\sigma_{b}(T)\sigma_{f}(T)}{\sigma_{b}(T) + \sigma_{f}(T)}
\approx \sigma_{c}(T) , \eqa where $\sigma_{b}(T)$ is the holon
conductivity, much smaller than fermion contributions justifying
the last approximation.

One can ask the role of the spinon conductivity for any physical
response functions. Actually, it contributes to the physical
thermal conductivity given by the corresponding Ioffe-Larkin
composition rule \bqa && \frac{\kappa(T)}{T} \approx
\frac{\kappa_{c}(T)}{T} + \frac{\kappa_{f}(T)}{T} , \eqa where
$\kappa_{c,f}(T)$ are thermal conductivity of conduction electrons
and spinons, respectively, and holon contributions are also
neglected. Assuming that the Wiedemann-Franz law holds for each
fermion sector, proven to be correct at least in the one loop
approximation \cite{Kim_TR}, we find \bqa &&
\frac{\kappa_{t}(T)}{T} \approx \frac{\pi^{2}}{3} \Bigl(
\sigma_{c}(T) + \sigma_{f}(T) \Bigr)  , \eqa suggesting that the
Wiedemann-Franz law should be violated due to the presence of
additional entropy carriers, that is, spinons at the Kondo
breakdown QCP in the low temperature limit, i.e., \bqa L(T) \equiv
\frac{\kappa (T)}{T \sigma (T)} \approx L_{0} \Bigl( 1 +
\frac{\rho_{f}v_{F}^{f}}{\rho_{c}v_{F}^{c}} \Bigr)      \eqa with
$L_{0} = \pi^{2}/3$, the value of the Fermi liquid. This result
would be robust beyond our approximation because this expression
includes just density of states and velocity at the Fermi energy,
thus expected to be governed by a conservation law.

\section{Summary}

In this paper we developed the quantum Boltzman equation approach
for the Kondo breakdown QCP, which two bands for conduction
electrons and localized fermions are involved with, where
scattering with $z = 3$ critical hybridization fluctuations and $z
= 3$ gapless gauge bosons relaxes their dynamics. Our main problem
was to understand the role of vertex corrections in their
transport phenomena, crucial for the $T$-linear non-Fermi liquid
resistivity in the $z = 3$ critical theory.

Only one band of spinons is involved for scattering with gauge
fluctuations, and their associated vertex corrections are
introduced in the usual way, demonstrated in the U(1) gauge theory
of section 3-1. Our treatment for gauge fluctuations is different
from the previous study \cite{YB_QBE}, in respect that the
vertex-distribution function is not well defined owing to its
formal divergence associated with vertex corrections, but
cancelled in the physical conductivity through self-energy
corrections as it should be, consistent with the diagrammatic
approach \cite{YB_Vertex}, while divergent contributions are
thrown away in the vertex-distribution function of the previous
study, well defined. Of course, both approaches give the same
result.

On the other hand, two bands should be taken into account for
scattering with hybridization excitations, giving rise to coupled
quantum Boltzman equations. In the so called decoupling limit
where coupling effects are neglected for the Boltzman equation of
spinons while they are allowed for that of conduction electrons,
the vertex correction for conduction electrons associated with
hybridization fluctuations turns out to be irrelevant due to heavy
mass of spinons denoted by $\alpha \ll 1$. Results of the
diagrammatic approach are recovered from the quantum Boltzman
equation approach in the decoupling limit.

The next task is what happens beyond the decoupling limit. Our
preliminary analysis shows that vertex corrections seem to appear
in the scattering channel with hybridization fluctuations.
However, we do not find the corresponding diagram for such a
correction at present. It remains as an interesting future study.

\section{Acknowledgments}

This work was supported by the French National Grant ANR36ECCEZZZ.
K.-S. Kim was also supported by the National Research Foundation
of Korea (NRF) grant funded by the Korea government (MEST) (No.
2009-0074542).


\section*{References}

\begin{thebibliography}{10}
\bibitem{HFQCP} Gegenwart P, Si Q, and Steglich F 2008
{\it Nature Physics} {\bf 4}, 186; Lohneysen H v, Rosch A, Vojta
M, and Wolfle P 2007 {\it Rev. Mod. Phys.} {\bf 79} 1015.
\bibitem{TR} Custers J, Gegenwart P, Wilhelm H,
Neumaier K, Tokiwa Y, Trovarelli O, Geibel C, Steglich F, P\'epin
C, and Coleman P 2003 {\it Nature} {\bf 424} 524.
\bibitem{HMM} Moriya T and Kawabata A 1973 {\it J. Phys. Soc. Jpn.} {\bf
34} 639; Moriya T and Kawabata A 1973 {\it J. Phys. Soc. Jpn.}
{\bf 35} 669; Hertz J A 1976 {\it Phys. Rev.} B {\bf 14} 1165;
Millis A J 1993 {\it Phys. Rev.} B {\bf 48} 7183.
\bibitem{LGW_F_QPT_Nature} Custers J, Gegenwart P, Wilhelm H,
Neumaier K, Tokiwa Y, Trovarelli O, Geibel C, Steglich F, P\'epin
C, and Coleman P 2003 {\it Nature} {\bf 424} 524.
\bibitem{GR_Exp} Kuchler R, Oeschler N, Gegenwart P,
Cichorek T, Neumaier K, Tegus O, Geibel C, Mydosh J A, Steglich F,
Zhu L, and Si Q 2003 {\it Phys. Rev. Lett.} {\bf 91} 066405.
\bibitem{dHvA} Shishido H, Settai R, Harima H, and Onuki Y 2005
{\it J. Phys. Soc. Jpn.} {\bf 74} 1103.
\bibitem{Hall} Paschen S, Luhmann T, Wirth S, Gegenwart P,
Trovarelli O, Geibel C, Steglich F, Coleman P, and Si Q 2004 {\it
Nature} {\bf 432} 881.
\bibitem{INS_Local_AF} Schroder A, Aeppli G, Coldea R, Adams M,
Stockert O, Lohneysen H v, Bucher E, Ramazashvili R, and Coleman P
2000 {\it Nature} {\bf 407} 351.
\bibitem{DMFT} De Leo L,  MCivelli, and Kotliar G 2008 {\it
Phys. Rev. Lett.} {\bf 101} 256404.
\bibitem{Senthil_Vojta_Sachdev} Senthil T, Vojta M, and
Sachdev S 2004 {\it Phys. Rev.} B {\bf 69} 035111.
\bibitem{Pepin_KBQCP} P\'epin C 2007 {\it Phys. Rev. Lett.} {\bf 98} 206401;
P\'epin C 2008 {\it Phys. Rev.} B {\bf 77} 245129.
\bibitem{Paul_KBQCP} Paul I, P\'epin C, and Norman M R 2007 {\it
Phys. Rev. Lett.} {\bf 98} 026402; Paul I, P\'epin C, and Norman M
R 2008 {\it Phys. Rev.} B {\bf 78} 035109.
\bibitem{Kim_GR} Kim K S, Benlagra A, and P\'epin C 2008 {\it
Phys. Rev. Lett.} {\bf 101} 246403.
\bibitem{Kim_TR} Kim K S and P\'epin C 2009 {\it
Phys. Rev. Lett.} {\bf 102} 156404.
\bibitem{Nagaosa_Lee} Lee P A and Nagaosa N 1992 {\it Phys. Rev.} B
{\bf 46} 5621.
\bibitem{FMQCP} Rech J, P\'epin C, and Chubukov A V 2006 {\it Phys.
Rev.} B {\bf 74} 195126.
\bibitem{Nambu_Vertex} Nambu Y 1960 {\it Phys. Rev.} {\bf 117}
648.
\bibitem{YB_Vertex} Kim Y B, Furusaki A, Wen X G, and Lee P A 1994
{\it Phys. Rev.} B {\bf 50} 17917.
\bibitem{YB_QBE} Kim Y B, Lee P A, and Wen X G 1995 {\it Phys. Rev.} B
{\bf 52} 17275.
\bibitem{Nave_QBE} Nave C P and Lee P A 2007 {\it Phys. Rev.} B {\bf
76} 235124.
\bibitem{U1SR} Lee S S and Lee P A 2005 {\it Phys. Rev.
Lett.} {\bf 95} 036403.
\bibitem{ACL} Kaul R K,Kim Y B, Sachdev S, and
Senthil T 2008 {\it Nature Physics} {\bf 4} 28.
\bibitem{SF_Kim} Kim K S and Kim M D 2008 {\it
Phys. Rev.} B {\bf 77} 125103; Kim K S and Kim M D 2007 {\it Phys.
Rev.} B {\bf 75} 035117.
\bibitem{Mahan_QBE} Mahan G D 2000 \textit{Many-Particle Physics} 3th
ed. (Kluwer Academic/Plenum Publishers, New York).
\bibitem{Hall_Conductivity} Thouless D J, Kohmoto M,
Nightingale M P, and den Nijs M 1982 {\it Phys. Rev. Lett.} {\bf
49} 405.
\bibitem{IF_Cond} Ioffe L B and Larkin A I 1989 {\it Phys. Rev.} B {\bf
39} 8988.
\end{thebibliography}
\end{document}